\documentclass[%
 reprint,
 amsmath,amssymb,
 aps,
]{revtex4-1}

\usepackage{graphicx}
\usepackage{dcolumn}
\usepackage{bm}
\usepackage{hyperref}
\usepackage[usenames,dvipsnames]{color}
\usepackage{subcaption}

\usepackage[normalem]{ulem}

\begin{document}


\title{Growth and Elasticity of Mechanically-Created Neurites}

\author{Madeleine Anthonisen}
 \email{madeleine.anthonisen@mail.mcgill.ca}
\author{Peter Gr\"utter}
\affiliation{%
Physics Department, McGill University, 3600 rue Universit\'e, Montr\'eal, Quebec, Canada H3A 2T8\\
}
 %

%
%

\date{\today}

\begin{abstract}

Working in the framework of morphoelasticity, we develop a model of neurite growth in response to elastic deformation. We decompose the applied stretch into an elastic component and a growth component, and adopt an observationally-motivated model for the growth law. We then compute the best-fit model parameters by fitting to force-extension curves from measurements of constant-speed uniaxial deformations of mechanically-induced neurites of rat hippocampal neurons. We find a time constant for the growth law of 0.009~s$^{-1}$, similar to the diffusion rate of actin in a cell. Our results characterize the kinematics of neurite growth and establish new limits on the growth rate of neurites.

\begin{description}
 \item[PACS numbers] 
\end{description}
\end{abstract}

\pacs{Valid PACS appear here}
\maketitle


%
%
%
%
\section{\label{sec:level1}Introduction}

Neurons are cells specialized for information processing. They have long, tube-like extensions of diameter $\sim1~\mu$m, termed neurites, that connect the cell bodies to other neurons and enable the exchange of information via chemical and electrical signals. Neurites are classified as axons, signal transmitters, or dendrites, signal receivers. 

\begin{figure}
\includegraphics[height=2.5in]{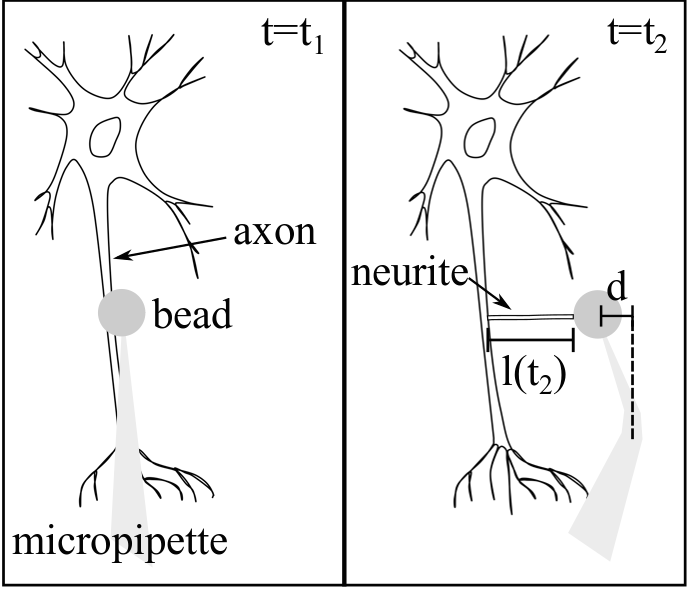}
\caption{\label{ns} Sketch of a neurite pulling experiment. At time $t_1$, the micropipette tip is fixed to the bead that contacts the axon of a neuron. At time $t_2$, the micropipette-bead complex has been moved by an amount $v(t_2-t_1)$ relative to the axon, inducing the growth of the neurite. The tension in the neurite is captured by recording the deflection of the micropipette, $d$, and calibrating its stiffness constant to convert the bending into a force.
}
\end{figure}

 Mechanical elongation of neurites has been widely studied (see e.g. \cite{franze2013mechanics,suter2011emerging} for reviews). These experiments have lead to the identification of tension as a driver of neurite growth and development \cite{athamneh2015quantifying,heidemann2015tension,o2008physical,zheng1991tensile}; e.g.~, ``a pulled axon grows as though  the nerve cell contained telescopic machinery prefabricated for elongation'' \cite{heidemann2015tension}. Recent work, \cite{suarez2013dynamics,magdesian2016rapid,ANTHONISEN2019121}, has shown that this telescopic growth also occurs in axon-like structures initiated from parent axons or dendrites, see Fig.~\ref{ns}. However, the mechanisms responsible for this surprising mass-accretion and the role of tension in limiting this process remain outstanding mysteries \cite{heidemann2015tension,athamneh2015quantifying,holland2015emerging,suter2011emerging}.
 
A natural question is the extent to which elongation can be attributed  to growth,  i.e. the addition of new cellular material, versus elastic stretching of existing constituents. In this paper, we answer this question.   

Working in the theoretical framework of morphoelasticity described in \cite{goriely2015neuromechanics,goriely2011morphoelasticity,goriely2017mathematics}, we relate the experimental force-extension curves of neurites to the material parameters  that describe their elastic response to deformations and the rate as well as the rates of material added due to growth.

In our experiments, we measure the force-extension relationship of new neurites using flexible, calibrated glass micropipettes as illustrated in Fig~\ref{ns}. The micropipette is connected to the cell by a bead that is chemically functionalized to induce a stable mechanical contact with the parent axon or dendrite. When the bead-pipette complex is displaced relative to the cell, the growth of an auxiliary structure, a new neurite, is induced. We elongate the neurite while simultaneously measuring its tension by optically tracking the beaded tip. By calibrating the spring constant of the pipette, we can convert this deflection to a force. We extend our neurites at a constant rate, in contrast with other experiments, e.g. \cite{bernal2007mechanical,dennerll1989cytomechanics}, in which a stretch is applied in one step and maintained constant for the duration of the experiment.


We derive an expression for the force-extension relationship of neurites that incorporates an exponential growth law. We fit experimental data to find the time constant for exponential mass addition, which is close to the rate of actin diffusion along a pulled neurite. We find that the time constants for different pulling experiments are positively skewed and follow a lognormal distribution. This puts new limits on the mass accretion of axon-like extensions. 

The structure of this paper is as follows: In Section~\ref{level3} we review the principles of morphoelastic theory and introduce a model to characterize the kinematics of neurite growth. We show the contributions of elastic stretching and growth stretching to neurite deformations in Section~\ref{level5}. In Section~\ref{level2} we justify assumptions used in \ref{level3} with experiments, summarizing this paper in Section~\ref{conc}.

\section{\label{level3}A model of growth with elastic deformation}


A general deformation can be characterized by a geometric stretch $\lambda$, defined as the relative change in the length of the neurite to the initial length, i.e. $\lambda\equiv l/L$, with  $l=l(t)$ and $L=l(t=0)$ the length of the neurite at time $t$ and the initial length respectively. 


We work within the framework of morphoelasticity, in which the geometric stretch is the product of an elastic term $\lambda^{e}$ and a growth term $\lambda^{g}$ \cite{goriely2015neuromechanics,goriely2011morphoelasticity,holland2015emerging,goriely2017mathematics}:
\begin{equation}
\lambda=\lambda^{e}\lambda^{g}.
\label{lambda}
\end{equation}
We assume that stress, defined as the axial force per unit area of the neurite cross-section, is only caused by elastic deformation, a commonly-adopted assumption in growth theories \cite{goriely2011morphoelasticity,goriely2015neuromechanics}. We further assume that the elastic part of the deformation is incompressible (volume preserving) so that only the growth component will add volume. 

In continuum mechanics, the stress-stretch relationship of soft materials is determined experimentally and can be derived from the strain energy density function of the deformation process. There are many different models to describe the strain energy density, we find that neurites are best described by the so-called Mooney-Rivlin model (this choice is justified in Sec.~\ref{level6}). Under the assumption of constant volume from the elastic deformation, the Mooney-Rivlin model has the form \cite{goriely2015neuromechanics}
\begin{equation}
\Psi=c_{1}[I_1^{e}-3]+c_2[I_2^{e}-3],
\label{mrim}
\end{equation}
where $c_1$, $c_2$ are material constants and $I^{e}_{1}=\lambda^{e2}_1+\lambda^{e2}_2+\lambda^{e2}_3$, $I^{e}_{2}=\lambda^{e2}_1\lambda^{e2}_2+\lambda^{e2}_2\lambda^{e2}_3+\lambda^{e2}_1\lambda^{e2}_3$ and $I^{e}_{3}=\lambda_1\lambda_2\lambda_3=1$ are elastic invariants in terms of the elastic stretches. Note $i=1,2,3$ label the spatial dimensions of the deformation. 
In the case of incompressible uniaxial extension, the neurite is pulled along a single dimension, $\lambda_1^{e}=\lambda^{e}$ and $\lambda_2^{e}=\lambda_3^{e}=1/(\lambda^{e})^{1/2}$. 
%
%

We can re-write the Mooney-Rivlin strain energy density function in terms of the elastic stretch and then reparameterize it in terms of $\lambda, \lambda^{g}$ \cite{goriely2015neuromechanics}:
\begin{equation}
\Psi(\lambda, \lambda^{g})=c_{1}\left[\left(\frac{\lambda}{\lambda^{g}}\right)^{2}+2\frac{\lambda^{g}}{\lambda}-3\right]+c_2\left[2\frac{\lambda}{\lambda^{g}}+\left(\frac{\lambda^{g}}{\lambda}\right)^{2}-3\right].
\label{psi}
\end{equation}
From Eq.~\ref{psi}, we can obtain the elastic Piola stress $P^e$, which can be used to obtain the Piola stress $P$, defined as \cite{goriely2015neuromechanics}:
\begin{equation}
P\equiv\frac{\partial\Psi}{\partial \lambda}.
\end{equation}
This can be expanded in terms of $\lambda$ and $\lambda^g$ as \cite{goriely2015neuromechanics},
\begin{equation}
P=\frac{2}{\lambda^g}\left[c_1+c_2\frac{\lambda^g}{\lambda}\right]\left[\frac{\lambda}{\lambda^g}-\left(\frac{\lambda^g}{\lambda}\right)^2\right].
\label{piola}
\end{equation}
The Piola stress captures the stress across the neurite. It can be related directly to an external loading force $F$ on a neurite through the principle of virtual work \cite{goriely2015neuromechanics,bower2009applied} to give%
%
\begin{equation}
F=PA,
\label{virt}
\end{equation}
where $A$ is the cross-sectional area.
%

The radial dimension of the neurite is a proxy for growth through addition of new material \cite{goriely2015neuromechanics}. In the absence of radial thickening, the transverse stretch $\lambda^{\bot}$, that is the ratio between the neurite radius at a time $t$, $r(t)$, and the initial radius $R$, is defined via the elastic stretch,
\begin{equation}
\lambda^{\bot}=\left(\frac{1}{\lambda^e}\right)^{1/2}.
\label{tran}
\end{equation}
The cross sectional area of a neurite can also be written in terms of $\lambda, \lambda^{g}$:
\begin{equation}
A=\pi R^2\left(\frac{\lambda^{g}}{\lambda}\right). 
\label{area}
\end{equation}
This allows the force $F$ to be written in terms of $\lambda$ and $\lambda^g$, and the parameters $c_1$ and $c_2$, as 
\begin{equation}
F=\frac{2\pi R^2}{\lambda} \left[c_1+c_2\frac{\lambda^g}{\lambda}\right]\left[\frac{\lambda}{\lambda^g}-\left(\frac{\lambda^g}{\lambda}\right)^2\right] .
\label{mothermother}
\end{equation}
From this one can compute not only the force at a given deformation, but also the full time evolution $F(t)$.

Indeed, axons under axial tension will gradually increase in mass to recover some homeostatic equilibrium state, that is the axon has been observed to have some inherent tension \cite{dennerll1989cytomechanics,lamoureux2010growth,goriely2015neuromechanics,goriely2011morphoelasticity}. Motivated by these observations, here we adopt a growth model in which the growth rate depends on the axial stress of the neurite. If the neurite is perturbed from that state, mass will be added so it can recover a particular ``baseline'' stress. To model growth, we assume a functional form of $\lambda^g$ based on experimental observations. 

%

\subsection{\label{beep} An exponential growth law}
Here we consider a growth law that states exponential growth or resorption occurs until a homeostatic stress is recovered. This model has been used to describe axonal growth in \cite{goriely2017mathematics}. Work from our lab indicates that the trajectory of actin (one of the principal constituents of neurites) entering the pulled neurite follows an exponential relation \cite{rigby2019building}.

We consider a law of the form, 
\begin{equation}
\frac{\partial\lambda^g}{\partial t}=k\lambda^g\left(\lambda^e-\lambda^* \right) \Theta(\lambda^e-\lambda^*),
\label{mod2}
\end{equation}
where $k$ is a constant, $\lambda^*$ is a critical stretch associated with the homeostatic stress $ \sigma^*$ that the neurite is trying to recover, and $\Theta(x)$ is a Heaviside theta function: $\Theta(x)$ is $1$ for $x>0$ and $0$ otherwise.

For $\lambda^e>\lambda^*$, Eq.~\ref{mod2} is 
\begin{equation}
\frac{\partial\lambda^g}{\partial t}=k(\lambda-\lambda^g\lambda^*).
\label{mod2s}
\end{equation}
Solving for the functional form of $\lambda^g(t)$ with the initial condition $\lambda^g(0)=1$, we obtain
\begin{equation}
\lambda^g(t)=\frac{k(L+vt)+v(e^{-kt}-1)}{kL},
\label{lamt2}
\end{equation}
where we have used $\lambda=(L+vt)/L$, $v$ is the (constant) speed at which neurites are extended. We set $\lambda^*=1$, an assumption we justify in a later section. 

\begin{figure}
\includegraphics[height=2.5in]{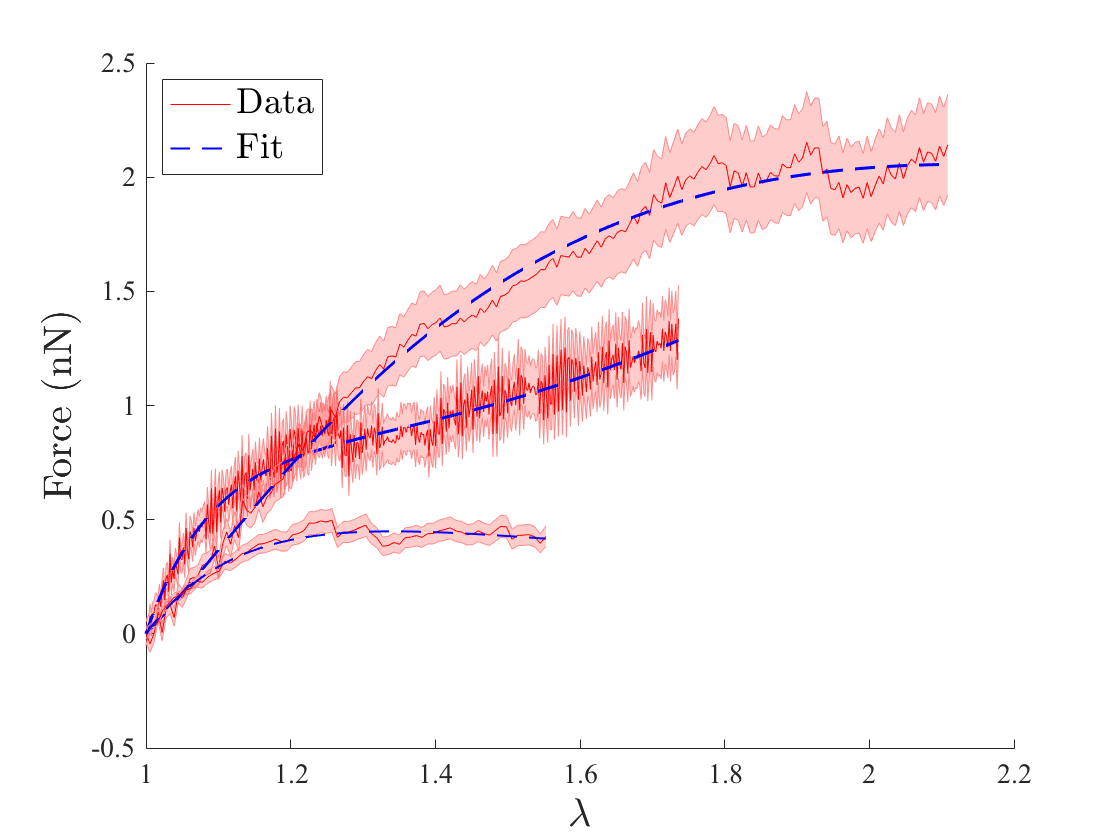}
\caption{\label{fits} Examples of force-extension curves of induced neurites. Each curve (red) is from a single pulling experiment and can be due to 1 or more mechanically-induced neurites. The force is the loading force applied axially to the neurite(s) as measured by the bending of the micropipette and the stretch $\lambda$ is the length the neurites have been pulled relative to their initial length. The shaded regions of the curves represent the measurement error and are calculated as described in \cite{ANTHONISEN2019121}. The dashed blue lines are fits of Eq.~\ref{mothermother}.
}
\end{figure}

In what follows, we will experimentally measure force-extension curves, and from this obtain the best-fit values of the  parameters $k$, $c_1$ and $c_2$ . Example experimentally-obtained force-extension curves are shown in Fig.~\ref{fits}.  Here we have fit the curves to the functional form Eq.~\ref{mothermother} with Eq.~\ref{lamt2} inserted.


\subsection{\label{level4}Material parameters}
\begin{figure}[ht!]
\begin{subfigure}[b]{0.5\textwidth}
   \includegraphics[height=2.7in]{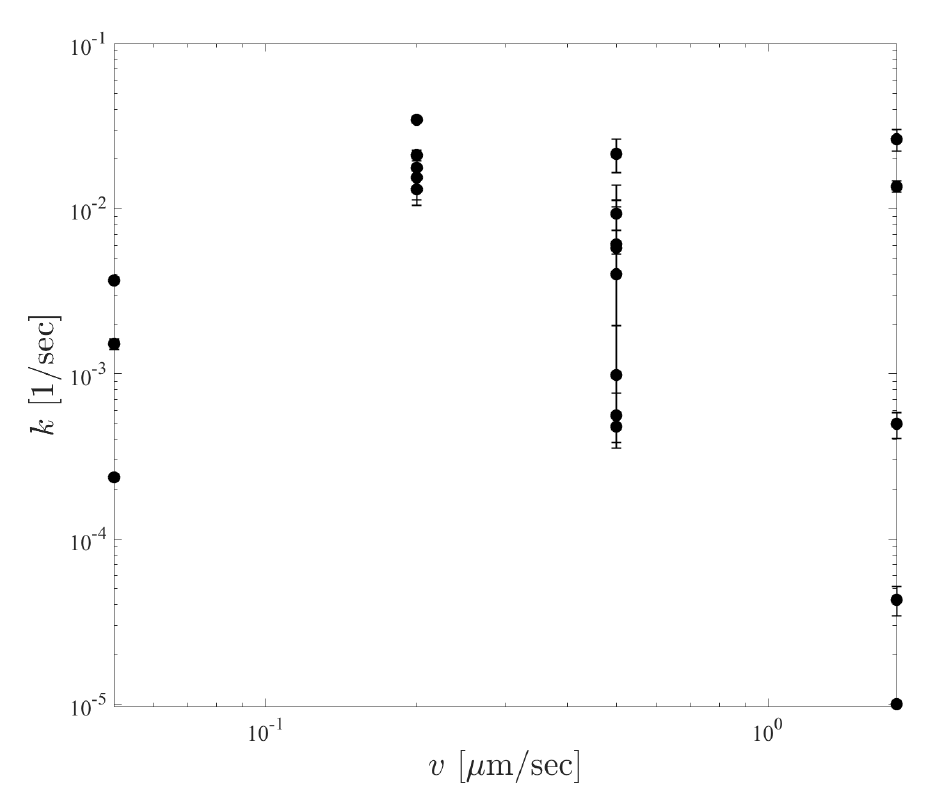}
   \caption{}
   \label{ks} 
\end{subfigure}

\begin{subfigure}[b]{0.5\textwidth}
   \includegraphics[height=2.48in]{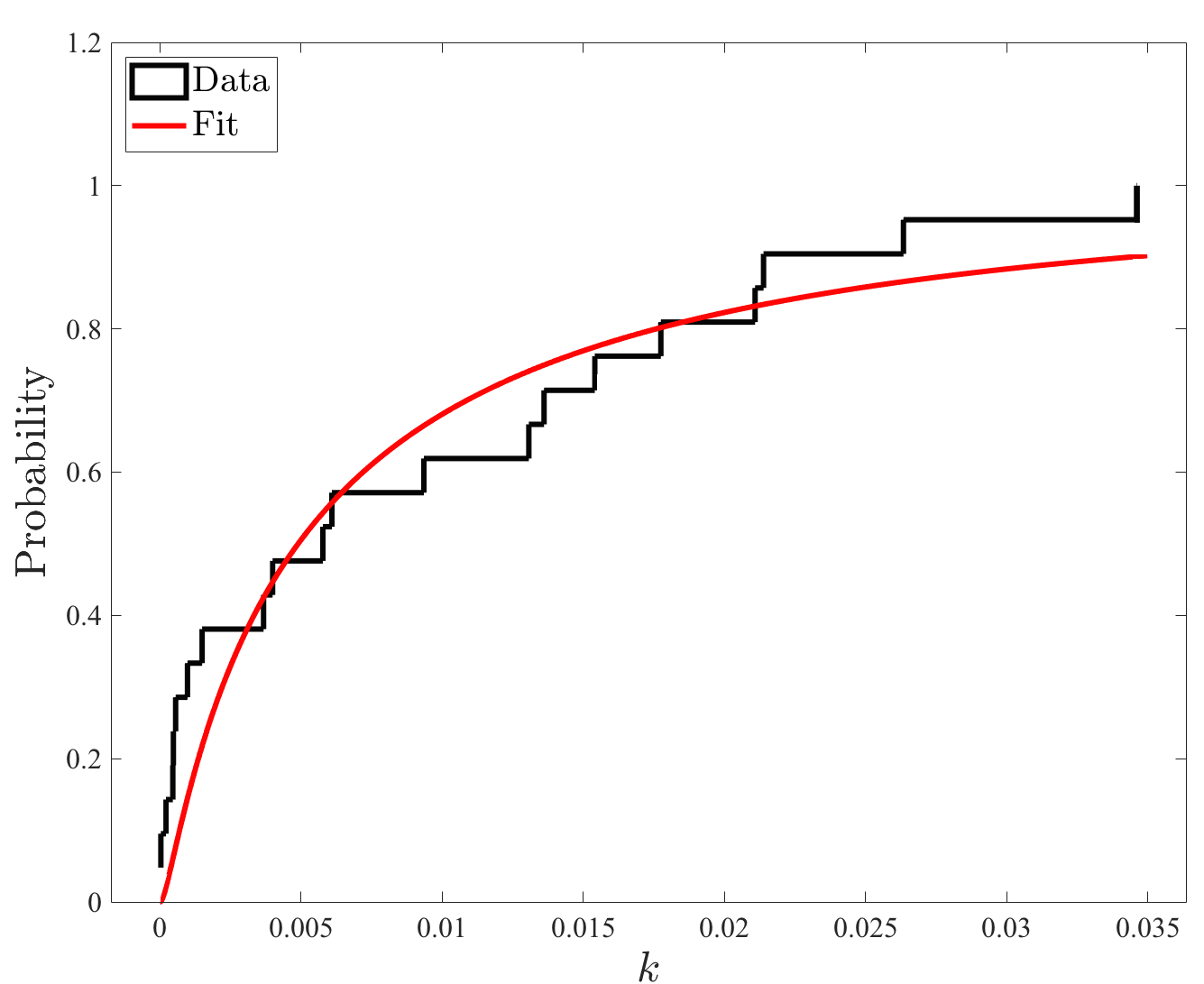}
   \caption{}
   \label{khist}
\end{subfigure}
\centering
\caption{\textbf{(a)} Time constants characterizing neurite growth plotted versus pull speed. This is the fit parameter $k$ from Eqs.~\ref{mothermother}\&\ref{lamt2}. The Kruskal-Wallis test confirms the hypothesis that $k$ values for each pull speed were all drawn from the same distribution. The mean value of $k$ across pull speeds is $0.009$ with a standard deviation of $0.01$. \textbf{(b)} Cumulative density function for $k$ values (black line), fit with a lognormal distribution (red line) that captures the skewness of the data. Parameters for the lognormal distribution are $\mu=-5.31$ and $\sigma=1.52$.}
\label{three}
\end{figure}

We fit twenty-one experimentally-obtained force-extension curves with Eq.~\ref{mothermother} and Eq.~\ref{lamt2} as described in Section~\ref{beep}. The growth rate parameter $k$ is shown in Fig.~\ref{ks} for different pull speeds. We find the mean value of $k$ to be $0.009$~s$^{-1}$, see Table~\ref{table1}. The data is skewed to large values of $k$, with a SD of $0.01$~s$^{-1}$. The mean is of a similar order of magnitude as the time constant describing the movement of actin along pulled neurites found in \cite{rigby2019building}, which is $0.001$~s$^{-1}$. The lower bound of our data matches the value found in \cite{holland2015emerging} for the axonal growth rate, 2$\times10^{-5}$~s$^{-1}$. In Fig.~\ref{khist}, we plot the cumulative density function of $k$ values and show that it is well-characterized by a lognormal distribution with parameters $\mu=-5.31\pm0.01$ and $\sigma=1.52\pm0.01$ (standard errors from fit). This is confirmed by a Chi-Square goodness of fit test at the 5$\%$ significance level. Here $\mu$ and $\sigma$ are the mean and standard deviation of the natural logarithm of $k$. The skewness, which captures the asymmetry of the distribution, can be obtained from $\sigma$ and is 33. Although this is significantly higher than 0.9, which is the sample skewness obtained from the definition of Pearson's moment coefficient of skewness, the two measures of skewness are consistent in describing the data as moderately to highly skewed.

We investigate the mass addition of new neurites as they are pulled, and find that $k$ is independent of pull speed. This is confirmed by the Kruskal-Wallis test, which tests whether samples, grouped by pull speed, are drawn from the same distribution. This isconsistent with previous work, \cite{ANTHONISEN2019121,anthonisen2019ps}, which found that neither the force-extension relationships nor cross-sectional areas of neurites depend on mechanical pull speed. This is surprising given a greater than 10-fold increase in pull speed. Interpreting $k$ as an exponential growth rate, it is reasonable that it should be the same across pull speeds as it could be constrained either by the properties of the cell (the speed with which it can manufacture and transport certain constituents) or by physical properties such as diffusion.

An open question is what causes the large variance in the $k$-values we extract, assuming this is not a feature that vanishes with more statistics. We postulate that this is related to the mechanisms underlying mass addition. Cell growth is typically modelled as a combination of active and passive processes. As mentioned, our timescales are consistent with those reported for actin in \cite{rigby2019building}. Motivated by this,
we consider the behaviour of actin filaments and myosin motors (the proteins responsible for polymerizing actin filaments) in a simplistic 1D diffusion model. We calculate the time it takes actin filaments and myosin motors, which together have an effective diffusion coefficient of $D=0.01~\mu$m$^2$s$^{-1}$ \cite{hannezo2015cortical}, to diffuse along a length equivalent to $L$ for each neurite. We find a mean rate of actin diffusion of $0.01$~s$^{-1}$ across neurites with a standard deviation of $0.01$~s$^{-1}$ that exactly matches our results for $k$. This is suggestive of an important role for actin diffusion in the addition of mass to new neurites. It also indicates that the variance in our reported values of $k$ could be explained by different initial neurite lengths.
%
%
%
%
%

The other material parameters, which characterize the neurite response to elastic stretch in the Mooney-Rivlin model, are $c_1=204\pm385$~kPa and $c_2=-13\pm302$~kPa (mean$\pm$SD). These are summarized in Table~\ref{table1}. Unlike $k$, the physical reason these parameters vary over orders of magnitude and, in the case of $c_2$, by sign is unclear. The Mooney-Rivlin model and the Neo-Hookean model, which is a specific case of the MR Eq.~\ref{mrim} \cite{de2016constitutive}, have been widely used to model other types of brain tissue \cite{bilston2001large,hrapko2006mechanical,rashid2013mechanical}, including axons \cite{holland2015emerging}. While this family of models is succesful in describing brain tissue under diverse experimental conditions, these other works also contain the feature that the material parameters are phenomenological and vary over orders of magnitude \cite{goriely2015neuromechanics}. In \cite{mihai2015hyperelastic}, it is shown analytically that the Mooney-Rivlin model, applied with different relative parameter signs, captures experimental trends observed in soft biological tissues under both shear and compression conditions. Our results add to the experimental evidence that the Mooney-Rivlin model is suitable to describe brain tissue. This indicates that the mechanical behaviour of newly induced neurites is very similar to that of naturally grown axons. While we lack a satisfying mechanistic interpretation of these parameters, quantifying single-cell behaviour with the Mooney-Rivlin model is an important step to multiscale modelling of the brain which could in turn ultimately clarify the physical significance of these results. 
\begin{table}
\caption{\label{table1}Fit parameters from Eqs.~\ref{mothermother}\&\ref{lamt2}.}
\begin{ruledtabular}
\begin{tabular}{ccccc}
 Parameter&Mean&Standard deviation\\ \hline
 $k$&0.009~s$^{-1}$
 &0.01~s$^{-1}$\\
 $c_1$&204~kPa&385~kPa \\
 $c_2$&-13~kPa&302~kPa\\
  %
   %
  %
 %
   %
 
\end{tabular}
\end{ruledtabular}
\end{table}

\section{\label{level5}Neurite growth}
\begin{figure}[tb!]
\includegraphics[width=0.5\textwidth]{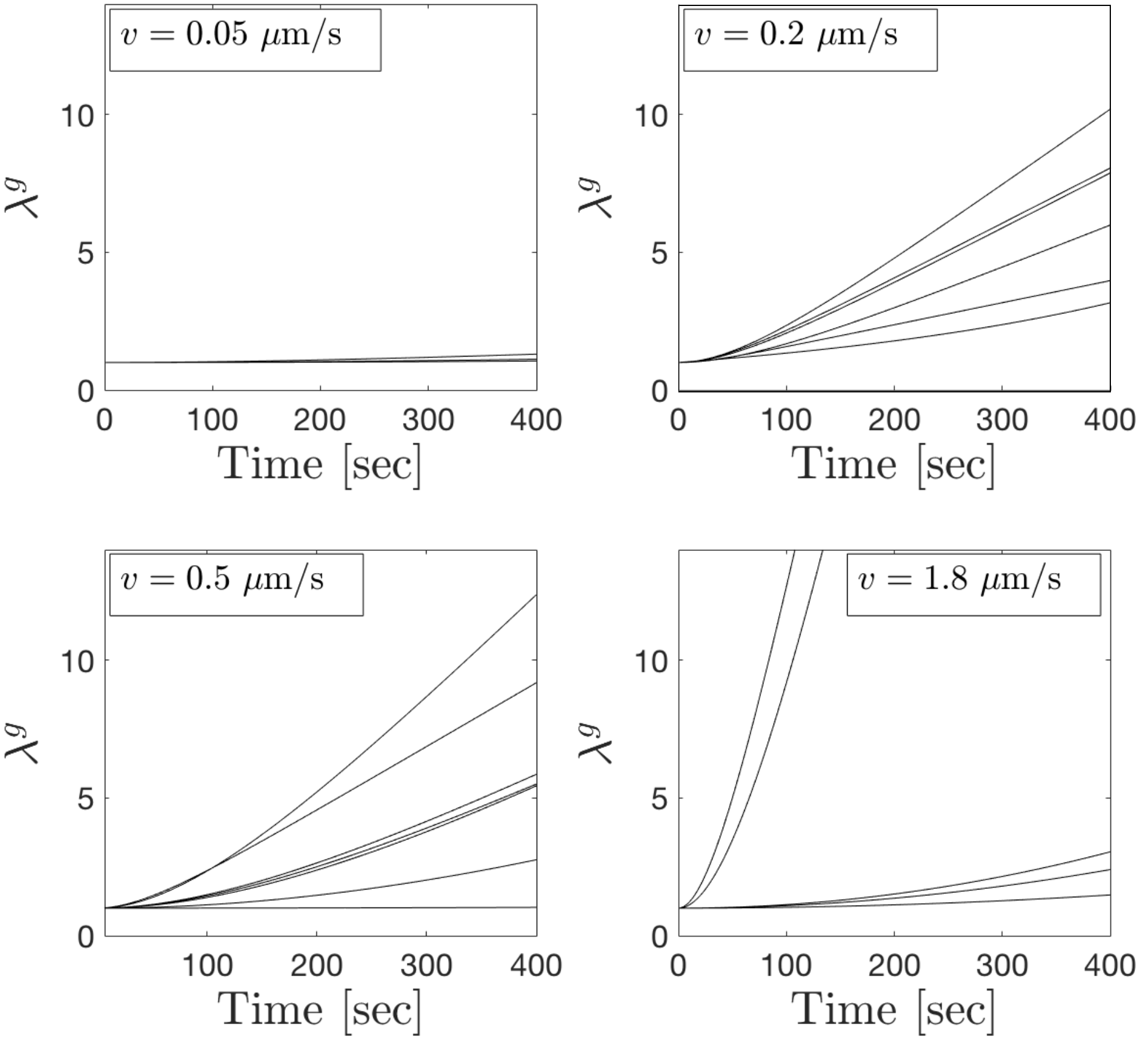}
\caption{\label{lambdat}Volume growth versus time for different pull speeds. In each plot, the differing curves correspond to differing initial neurite length.
}
\end{figure}

In Fig.~\ref{lambdat} we plot the component of neurite stretch that is due to added mass, $\lambda^{g}$, versus time. Each curve represents one pulling experiment and we plot Eq.~\ref{lamt2} with the $k$ corresponding to the force-extension curve. These curves show the rapidity with which new material is added to neurites as they are being pulled. They represent the volume growth of neurites. 

With the exception of $v=0.05~\mu$m/s, the $\lambda^{g}$ curves take on a range of values for a single pull speed. As discussed in Section~\ref{level4}, this could be due to the different initial lengths of neurites, which for $v=0.05~\mu$m/s were in the 52$^{\mathrm{nd}}$, 90$^{\mathrm{th}}$ and 95$^{\mathrm{th}}$ percentiles of the data respectively; this is consistent with the idea that mass addition is less extreme if the diffusion path for material along the neurite is longer.

\begin{figure}[!]
\includegraphics[width=.5\textwidth]{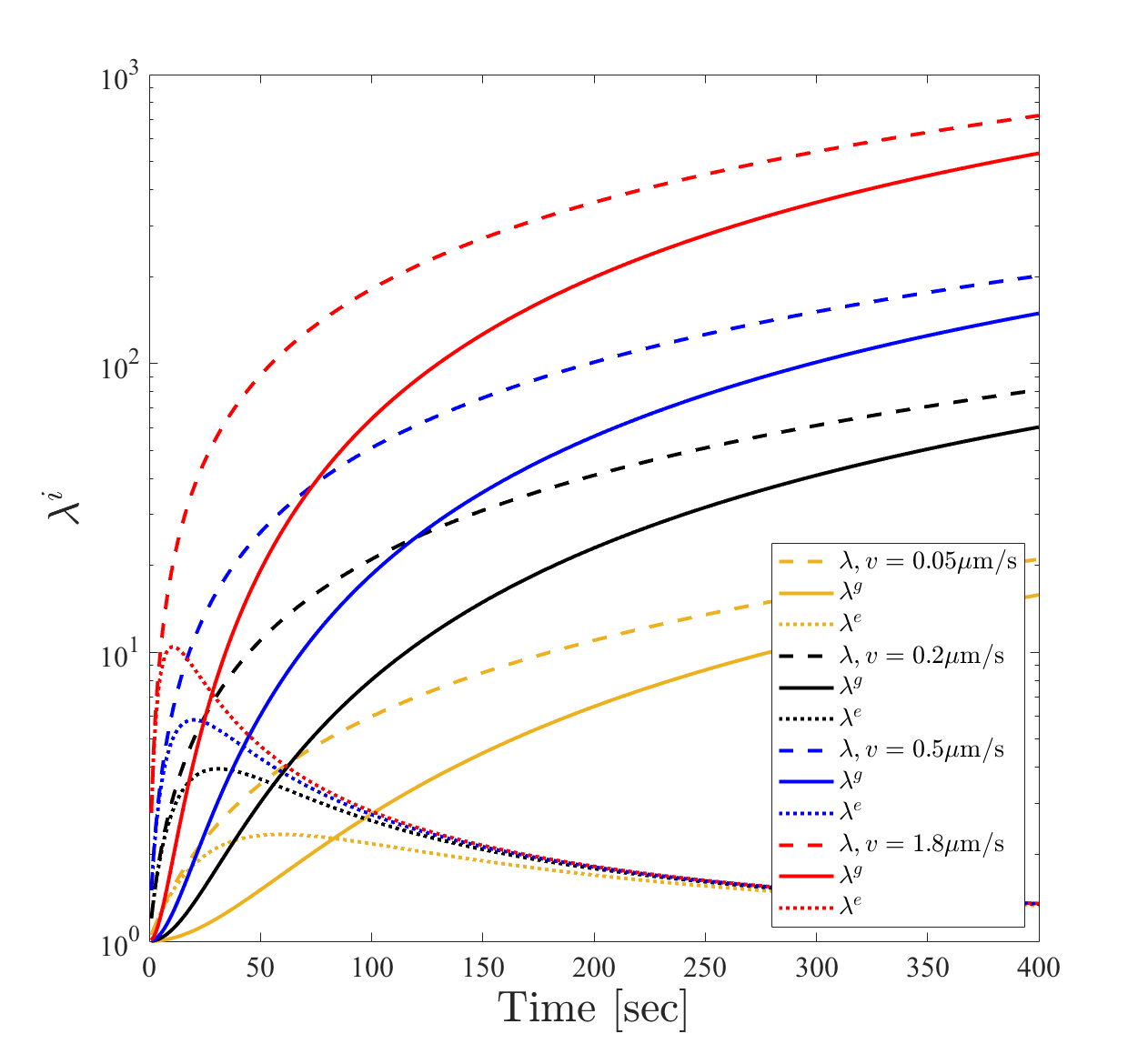}

\caption{\label{lambdas2}Stretch contributions of neurites. The different colours correspond to different pull speeds and the different line types (dashed, solid, dotted) are associated with different types of stretch ($\lambda$ is dashed, $\lambda^{g}$ is solid and $\lambda^{e}$ is dotted). $\lambda$ for each pull speed (dashed lines) is calculated with an initial neurite length of $L=1~\mu$m. The volume addition $\lambda^{g}$ (solid lines) for each pull speed are computed with Eq.~\ref{lamt2} taking the mean $k=0.009$~s$^{-1}$ from all pulling experiments. We see that these track $\lambda$, approaching it for later times. $\lambda^{e}=\lambda/\lambda^{g}$ are plotted with dotted lines and show the elastic response of the neurite for different pull speeds. Initially all $\lambda^{e}$ track $\lambda$ then all collapse to values approaching 1.
}
\end{figure}

In Fig.~\ref{lambdas2}, we show the different contributions to neurite deformation, $\lambda$, from the elastic stretch $\lambda^{e}$ and the growth stretch $\lambda^{g}$. We see that for all the pull speeds, initially there is no growth and the entire deformation is elastic, $\lambda(0)=\lambda^{e}$. With time, the neurite grows according to Eq.~\ref{mod2s} and we see $\lambda^{g}\rightarrow\lambda$ and $\lambda^{e}\rightarrow1$. This reflects the the fact that the elastic stretch is evolving to recover a homeostatic equilibrium value. Interestingly, after $\sim$175~s, the elastic stretches for all speeds collapse to the same values approaching 1. This indicates that for the range of speeds studied, there is a point past which the elastic response is independent of pull speed. This timescale is associated with the mechanisms of mass addition: As the cell has more time to add mass to the neurite, the stretch response of the existing neurite-components becomes less significant. For each speed, there is a time at which $\lambda^{e}=\lambda^{g}$. This time is inversely proportional to the pull speed meaning neurites are very flexible in their responses to deformation and able to accommodate loading forces applied over a large range of speeds. 

We see from the form of Eq.~\ref{lamt2} that $\lambda^{g}$ tracks $\lambda$ with a pull-speed-dependant exponential term. We pull at extremely fast speeds relative to physiological ones and relative to other pulling experiments in the literature. Assuming our growth model, Eq.~\ref{lamt2}, is realistic, these results show new limits of mass addition for axon-like extensions.
%
%
%
%

\section{\label{level2}Results from experiments}
Before we conclude, we address and experimentally verify the validity of the assumptions made in this analysis. 

\vspace{-1cm}

\subsubsection{\label{level6}Model selection}

We have used the Mooney-Rivlin model to characterize the neurite response to deformation. However there do exist alternative choices.

\begin{figure}
\includegraphics[height=3.1in]{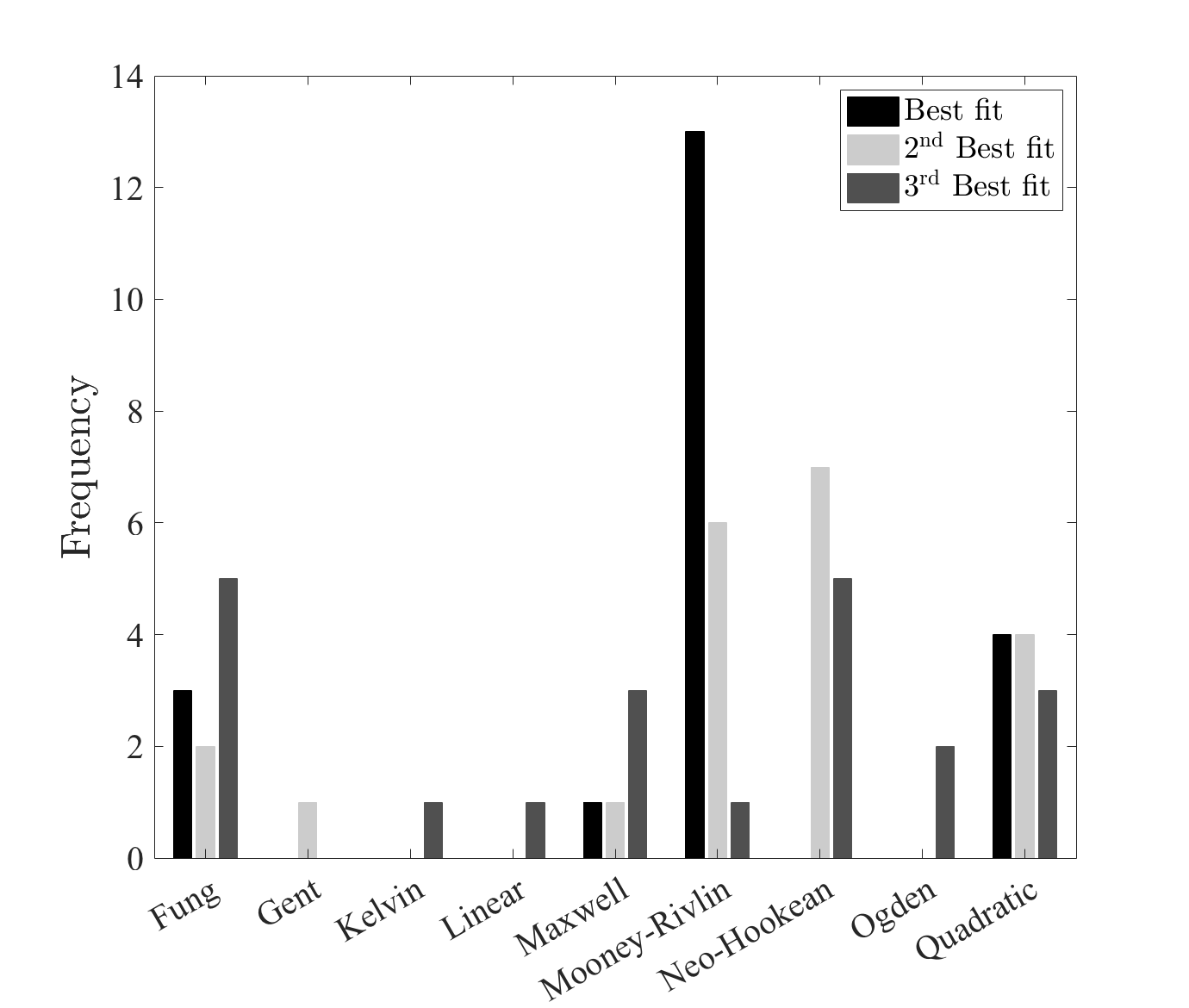}
\caption{\label{bf} Relationships that best fit the data as determined by minimizing the AIC. Of the models commonly used to characterize the stress-strain relationship, the Mooney-Rivlin model was most frequently the best fit.}
\end{figure}

To model the mechanical response of neurites to stretch, we compare a series of widely-used constitutive models, including viscoelatic and hyperelastic relations\cite{mondaini2008mathematical,goriely2015neuromechanics,de2016constitutive,goriely2017mathematics}. We obtain constitutive relationships from strain-energy density functions, $\Psi$, and fit these to each curve. We determine the best fit by minimizing the Akaike Information Criterion (AIC)\cite{wagenmakers2004aic,snipes2014model}. Fig.~\ref{bf} is a bar graph showing the frequency of `wins' of each constitutive relationship, that is the number of curves for which that relationship gave the minimal AIC. This demonstrates that a Mooney-Rivlin relation best describes the data. 

%

\subsubsection{\label{level7}Added volume}
The derivation of Eq.~\ref{virt} assumes $A$ is homogeneous along the axial length of the neurite \cite{goriely2015neuromechanics}. In \cite{anthonisen2019ps}, a method for extracting radii of neurites below the optical-diffraction limit is developed and it is shown that neurites have a constant radius along their length a short time after they are pulled. 


Together, radius measurements and our analysis confirm volume growth along the neurite. If volume were conserved during neurite deformations, then $\lambda^{\bot}=1/\lambda^{1/2}$ \cite{fan2017coupled}. In our framework, $\lambda\neq\lambda^{e}$ at later times so volume is not conserved. 

From the form of $\lambda^{e}$ (Fig.~\ref{lambdas2}), we see that initially $\lambda^{\bot}$ decreases to accommodate stretch since mass flow is limited on very short timescales. With time, $\lambda^{\bot}$ increases, tending towards 1. Ref.~\cite{lamoureux2010growth} reported radial thinning then thickening along the axon but on much longer timescales (several hours). We apply stress at much faster rates than \cite{lamoureux2010growth} and our neurites are on average $\sim5\times$ smaller than axons. These factors could potentially explain the faster mass accretion rates observed here. Faster rates of applied stress could trigger a faster response and thinner neurites can increase their relative mass more quickly.

\subsubsection{\label{level8}Critical stretch}
\begin{figure*}
\includegraphics[height=2.85in]{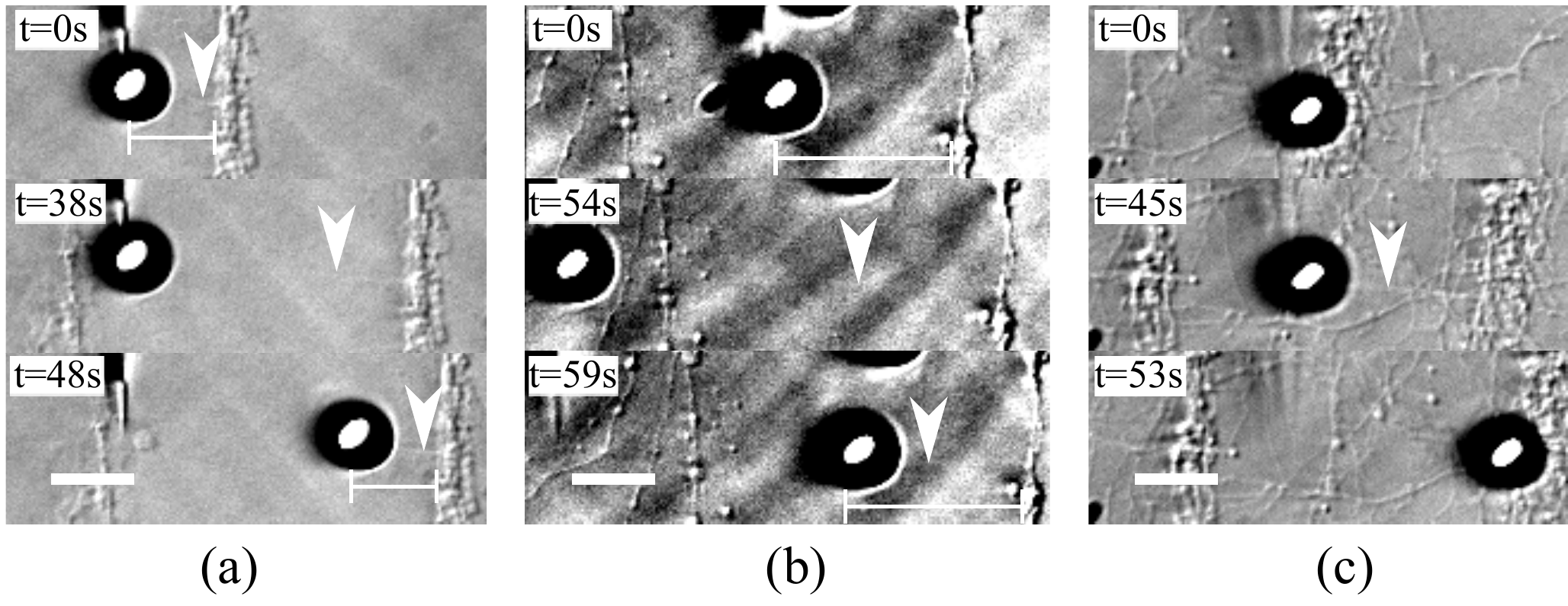}
\caption{\label{lf}\textbf{(a-c)} Snapshots of experiments where the beads detach from the pipettes and the neurites (white arrows) return to their initial lengths. Thick scale bars are $10~\mu$m. Thin scale bars represent different initial lengths; these are absent in \textbf{(c)} where the neurite is obscured by the bead. The second image in each series is the frame before the bead detaches.}
\end{figure*}

Here we present experimental evidence of a critical stretch, $\lambda^*$, associated with a critical stress, $\sigma^*$ that the neurite is trying to recover. Although we do not have sufficient statistics to concretely state $\lambda^*=1$, we have indication that this is a reasonable choice, see Fig.~\ref{lf}.

In our experiments, neurites are initiated by initially pulling the bead very slowly for $1-5~\mu$m, see first snapshot of Fig.~\ref{lf}c \cite{magdesian2017rewiring}. In some cases, the neurite was pulled even further ($\sim10-20~\mu$m, see Fig.~\ref{lf}a-b) to ensure that we could pull at greater rates without the bead becoming dislodged by other cells or debris on the coverslip. This slow initiation length was taken to be the initial length $L$ in the analysis described above. This is the starting point for rapid pulling.

In some instances, during pulls at high pull speeds ($0.05-1.8~\mu$m/s), the neurite was stronger than the suction applied to fix the bead to the tip of the micropipette. In these cases, the bead would return to its initial position---indicating that the neurite has a critical tension it is trying to restore. 

\section{\label{conc}Conclusion}
In this work we have developed a model that links tension in extending neurites to the rate of mass addition. This lets us quantify the role of tension as a driver of neurite growth. The fact that the mechanical behaviour of induced neurites is similar to naturally-grown axons under stretch indicates that our pulling experiments are relevant to questions of axon growth \cite{mondaini2008mathematical,goriely2015neuromechanics,de2016constitutive,goriely2017mathematics} . We quantify a new capacity for growth through the addition of new material.

Using a Mooney-Rivling model, we identify the contribution of hyper-elastic stretching to neurite deformation under loading. We find the material constants $c_1$ and $c_2$ vary over orders of magnitude without a satisfying reason as to why. However, we add a Mooney-Rivlin characterization of new structures to the existing body of literature. In future, this could be used in multi-cell models of the brain. 

Motivated by previously-reported observations  \cite{rigby2019building}, we adopt an exponential growth law to model mass addition. We find that the time constants $k$ are distributed lognormally. The mean value of $k$ is close to the time constant of diffusion of actin in neurites, which could indicate the importance of diffusion in the growth process.
\section*{\label{ack}Acknowledgements}
Both authors would like to acknowledge funding from the National Sciences and Research Council of Canada (NSERC) and the Fonds de recherche du Qu\'ebec---Nature et technologies (FRQNT). We fervently thank Evan McDonough for invaluable discussions and recommendations for analysis.
\bibliography{hmmm}
\end{document}